\documentclass{ws-p8-50x6-00-hack}

\newcommand{\bm}[1]{\mbox{\boldmath $#1$}}

\newcommand{\bkt}{\bm k_{T}}

\newcommand{\ba}{\begin{eqnarray}}
\newcommand{\ea}{\end{eqnarray}}
\newcommand{\beq}{\begin{equation}}
\newcommand{\eeq}{\end{equation}}

\newcommand{\text}[1]{{\rm #1}}

\begin{document}

\title{Transversity single spin asymmetries\footnote{
Talk presented at DIS2001, Bologna, April 27 - May 1, 2001}}

\author{\mbox{}\\[-6 mm]
Dani\"{e}l Boer}

\address{RIKEN-BNL Research Center, Brookhaven National Laboratory\\
Upton, NY 11973, USA\\ 
E-mail: dboer@bnl.gov}


\maketitle

\abstracts{The theoretical aspects of two leading twist 
transversity single spin asymmetries, one arising from the Collins effect
and one from the interference fragmentation functions, are reviewed. 
Issues of factorization, evolution and Sudakov factors for the relevant 
observables are discussed.
These theoretical considerations pinpoint the most realistic scenarios towards 
measurements of transversity.\mbox{}\\[-6 mm]} 

\section{Collins effect asymmetries}

\begin{figure}[htb]
\begin{center}
\leavevmode \epsfxsize=7.5cm \epsfbox{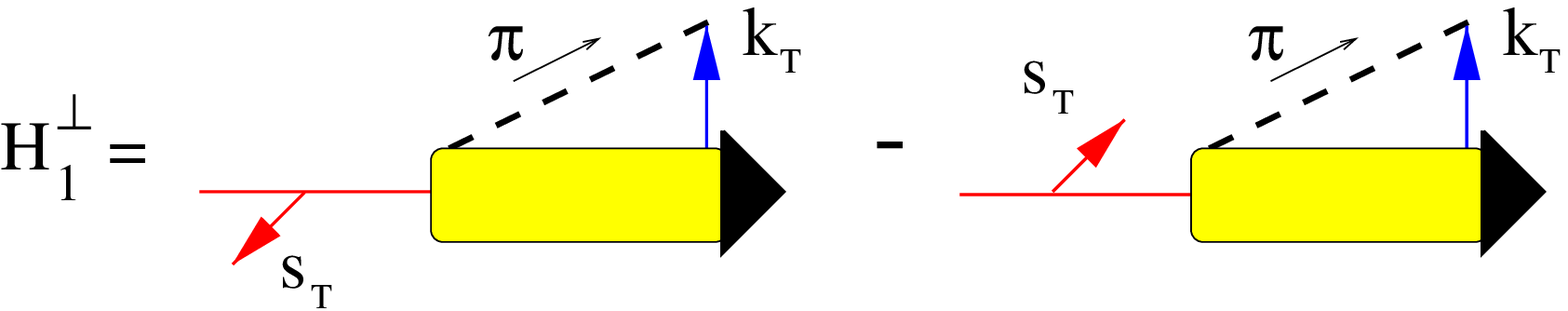}
\caption{\label{Collinseffect}The Collins effect: $q(\bm s_{T}) \to \pi (\bkt) 
X \, \neq \, q(-\bm s_{T}) \to \pi (\bkt) X $.}
\end{center}
\vspace{-2 mm}
\end{figure}

\noindent
The Collins effect refers to a nonzero correlation between 
the transverse spin $\bm s_{T}$ of a fragmenting quark and the distribution of 
produced hadrons. More specifically, a transversely polarized quark can 
fragment into particles (with nonzero transverse momentum $\bkt$) having a 
$\bkt \times \bm s_{T}$ angular distribution around the jet axis or, 
equivalently, the quark momentum, see Fig.\ \ref{Collinseffect}. 
The Collins
effect will be denoted by a fragmentation function $H_1^\perp(z,\bkt)$ and
is expected to be nonzero due to final state interactions 
between a measured final state hadron (e.g.\ a $\pi$) and the rest of the jet 
($X$). The Collins effect can lead to single spin asymmetries (SSA) in 
$e \, p^\uparrow \to e' \pi \, X$ and $p \, p^\uparrow \to \pi \, X$. 
There exist some experimental indications that the Collins effect is 
nonzero, e.g.\ SSA measured by HERMES \cite{HERMES,DB-99} 
and SMC \cite{Bravar} at relatively low energies.

\subsection{Collins effect in semi-inclusive DIS}

\begin{figure}[htb]
\begin{center}
\leavevmode \epsfxsize=7cm \epsfbox{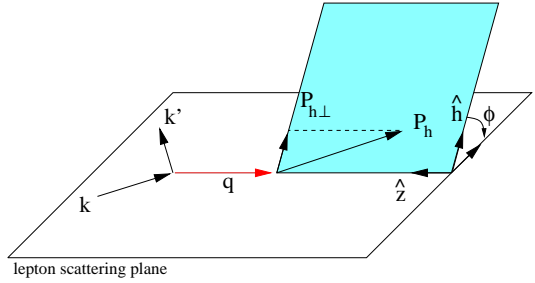}
\caption{\label{kindis}Kinematics of semi-inclusive DIS.}
\end{center}
\vspace{-2 mm}
\end{figure}

\noindent
Collins \cite{Collins-93} considered semi-inclusive DIS $e + p^\uparrow
\to e' + \pi + X$, where the spin of the proton is orthogonal to the 
direction of the virtual photon $\gamma^*$ and one observes the 
transverse momentum $\bm P^{\pi}_{\perp}$ of the $\pi$ in the jet, which has 
an angle $\phi^e_\pi$ compared to the lepton scattering plane 
($\phi$ in Fig.\ \ref{kindis}).
Collins has shown that the cross section for this process has an asymmetry 
that is proportional to the transversity function $A_{T} \propto  
\sin(\phi^e_\pi + \phi^e_S) \; |\bm S_{T}^{}| \; h_1 \; H_1^\perp$.
To discuss this SSA further, we will first project it out from the cross 
section (cf.\ Ref.\ \cite{Boer-Mulders-98}). Consider the cross sections 
integrated, but weighted with a function $W = W(|\bm
P^{\pi}_{\perp}|,\phi_{\pi}^e)$ of the transverse momentum of the $\pi$:  
\beq
\langle W \rangle
\equiv \int d^2\bm P^{\pi}_{\perp}
\ W\,\frac{d\sigma_{P_e P_p}^{[e \, p^\uparrow \rightarrow e' \,
\pi \, X]}} {dx\,dy\,dz\,d\phi^e\,d^2\bm P^{\pi}_{\perp}},
\eeq
where we restrict to the case of $|\bm P^{\pi}_{\perp}|^2 \ll Q^2$.  
We will focus on 
\beq
{\cal O} \equiv \frac{\big\langle 
\sin( \phi_{\pi}^e + \phi_{S}^e) \, |\bm P^{\pi}_{\perp}|  
\big\rangle}{{\scriptstyle 
\left[4\pi\,\alpha^2\,s/Q^4\right]}M_\pi} = 
\vert \bm S_{T}\vert\,{\scriptstyle (1-y)} 
\sum_{a,\bar a} e_a^2
\,x\,h_{1}^{a}(x) z H_1^{\perp (1) a}(z),
\label{observableA}
\eeq
where 
\beq
H_1^{\perp (1)}(z) \equiv \int 
d^2 \bkt \frac{\bkt^2}{2 z^2 M_\pi^2} H_1^{\perp}(z,\bkt^2).
\eeq
At present all phenomenological studies of the Collins effect are performed 
using such tree level expressions. 
On the other hand, the leading order (LO) evolution equations 
are known for $h_1$ \cite{Baldra} (NLO even) and $H_1^{\perp (1)}$ 
(at least in the large $N_c$ limit \cite{Henneman}). 
Both functions evolve autonomously and vanish asymptotically.  
The following question arises: 
if one measures ${\cal O}$ at different energies, can one relate them 
via LO evolution? The answer is: yes, the LO $Q^2$ 
corrections to the {\em tree level} observable ${\cal O}$ arise only from the 
evolution of $h_1$ and $H_1^{\perp (1)}$. This is a nontrivial result, 
since this semi-inclusive process is not a case where collinear factorization
applies. In the differential cross section $d\sigma/d^2\bm P^{\pi}_{\perp}$ 
itself, beyond tree level soft gluon corrections do not cancel; Sudakov 
factors need to be taken into account; a more complicated factorization 
theorem applies \cite{CS-81,DB-01}.

\begin{figure}[htb]
\begin{center}
\leavevmode \epsfxsize=7.5cm \epsfbox{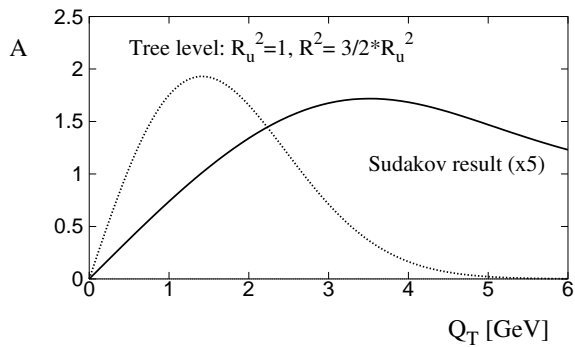}
\caption{\label{Col902}The asymmetry factor ${\cal A}(Q_T)$ (in units of
$M_h$). The solid curve is a generic Sudakov factor result  
(at $Q = 90 \, \text{GeV}$) multiplied by a factor 5. The other curve is the 
tree level quantity for certain Gaussian transverse momentum widths.}  
\end{center}
\vspace{-2 mm}
\end{figure}

Here we will briefly consider the effects of Sudakov factors in the explicit 
example of the Collins effect asymmetry in semi-inclusive DIS $e \, p \to e'
\, \gamma^*(\bm{q}_{T}^{}) \, p \to e' \, \pi \, X$ ($\bm{q}_{T}^{} = -z \bm
P^{\pi}_{\perp}$ and $\bm{q}_{T}^2 \equiv Q_T^2 \ll Q^2$)
\beq
\frac{d\sigma(e\, p \to e' \pi X)}{dx dz dy d\phi_e d^{\,2}{\bm
q_{T}^{}}}
\propto 
\left\{ 1 + |\bm S_{T}^{}|\;\sin(\phi^e_\pi+\phi^e_{S})\; 
A(\bm{q}_{T}^{}) \right\}.
\eeq
To get an idea about the
effect of Sudakov factors, we will assume Gaussian transverse 
momentum dependence for $H_1^\perp$. The asymmetry analyzing power is then 
given by 
\beq
A(\bm{q}_{T}^{}) = \frac{
\sum_{a}\;e_a^2 \; B(y)\; h_1^a(x) H_1^{\perp (1) a}(z)
}{\sum_{b}\; e_b^2 \; A(y)\;f_1^b(x) D_1^b(z)} {\cal
A}(Q_T),
\eeq
where $A(y)= (1-y+\frac{1}{2} y^2), B(y)=(1-y)$.
Furthermore, since the nonperturbative Sudakov 
factor ($S_{NP}$) is not determined from SIDIS experiments 
(despite the ZEUS data), 
for illustration purposes we will use the parameterization of 
Ladinsky-Yuan \cite{LY}. In Fig.\ 
\ref{Col902} ${\cal A}(Q_T)$ is given at $Q=M_Z$ and compared to the tree 
level result for Gaussian transverse
momentum widths chosen such as to minimize that result (values more typical
of a tree level analysis produce a larger asymmetry factor).
We refer to Ref.\ \cite{DB-01} for details.   
We see that ${\cal A}(Q_T)$ at $Q=M_Z$ becomes considerably smaller
($\max({\cal A}(Q_T)) \sim Q^{-0.5} - Q^{-0.6}$) and broader
than the tree level expectation.
Thus, tree level estimates tend to overestimate
transverse momentum dependent azimuthal spin asymmetries and 
Sudakov factors cannot be ignored at present-day collider energies.

\subsection{Collins effect in $e^+ \, e^- \to \pi^+ \, \pi^- \, X$} 

\begin{figure}[htb]
\begin{center}
\leavevmode \epsfxsize=7.5cm \epsfbox{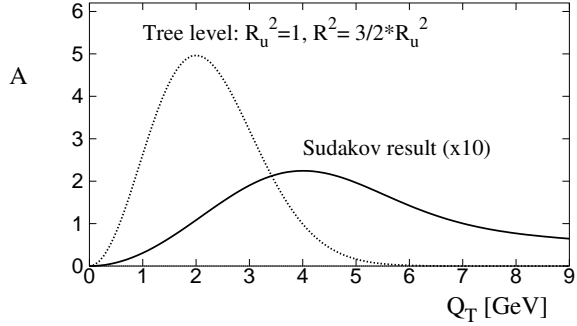}
\caption{\label{Cos902}The asymmetry factor ${\cal A}(Q_T)$ (in units of
$M^2$). The solid curve is a generic Sudakov factor result  
(at $Q = 90 \, \text{GeV}$) multiplied by a factor 10.
The other curve is the tree level quantity for certain Gaussian transverse
momentum widths.}  
\end{center}
\vspace{-2 mm}
\end{figure}

In order to obtain the Collins function, one can measure a $\cos(2\phi)$
asymmetry in $e^+ \, e^- \to \pi^+ \, \pi^- \, X$, that is essentially 
proportional to the Collins function squared \cite{BJM-97} (at average
momentum fractions). 
\begin{figure}[htb]
\begin{center}
\leavevmode \epsfxsize=7.5cm \epsfbox{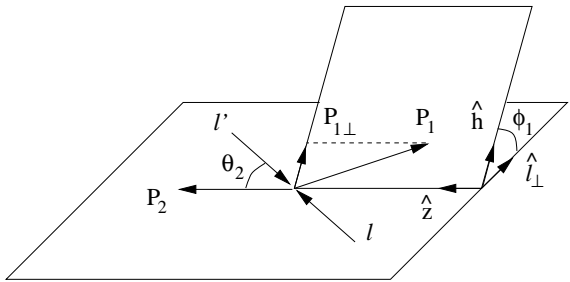}
\caption{\label{fig:kinann} Kinematics of $l \, l' \to P_1 \, P_2 \, X$, where
$l, l'$ are the momenta of $e^-, e^+$; 
$P_1, P_2$ are the momenta of hadrons in opposite jets.}
\end{center}
\vspace{-2 mm}
\end{figure}
A first indication of such a nonzero asymmetry comes from a 
preliminary analysis \cite{EST} of the 91-95 LEP1 data ({\small
DELPHI}). Also, the possibilities at BELLE  
are currently being examined \cite{Perdekamp}.
The extraction of the Collins function from this asymmetry is not
straightforward, since there is asymmetric background from hard gluon 
radiation (with $Q_T \sim Q$) and from weak decays. Moreover, using the tree 
level 
asymmetry expression is not sufficient and beyond tree level Sudakov factors 
need to be included. If the differential cross section is written as 
\beq
\frac{d\sigma (e^+e^-\to \pi^+ \pi^- X)}{d\Omega dz_1 dz_2 d^2{\bm
q_{T}^{}}} 
\propto \left\{ 1 + \cos(2\phi_1) A(\bm{q}_{T}^{}) \right\},
\eeq
with $\bm{q}_{T}^2 \ll Q^2$, then assuming again Gaussian
transverse momentum dependence for the Collins function, we find
\beq
A(\bm{q}_{T}^{}) = \frac{
\sum_{a}\;c_2^a\;B(y)\;H_1^{\perp (1) a}(z_1) \; \overline H{}_1^{\perp
(1) a}(z_2)}{\sum_{a}\; c_1^a\;A(y)
\; D_1^a(z_1) \; \overline D{}_1^a(z_2) }\; {\cal A}(Q_T), 
\eeq
with somewhat different $A(y), B(y)$. 
Surprisingly there is also no determination of $S_{NP}(b)$ from $e^+ \; e^- \to
A \; B \; X$, so for illustration purposes we use again 
Ladinsky-Yuan's $S_{NP}(b)$. The result is displayed in Fig.\ \ref{Cos902} and
compared to a conservative (i.e.\ expected to be too small) 
tree level curve.
This generic example \cite{DB-01} shows that Sudakov factors   
produce an order of magnitude suppression at $Q=M_Z$ ($\max({\cal A}(Q_T)) 
\sim Q^{-0.9} - Q^{-1.0}$), hence this Collins 
effect observable is best studied with two jets events at $\sqrt{s} \ll M_Z$. 

\section{Interference fragmentation functions}

\noindent
Jaffe, Jin and Tang \cite{JJT} pointed out the possibility that the Collins 
effect averages to zero in the sum over final states $X$. 
Instead, they proposed to measure two pions in the final state
$\vert (\pi^+ \, \pi^-)_{\text{out}} X \rangle $ ($\pi^+, \pi^-$ 
belong to the same jet), which presumably depends on the strong phase 
shifts of the $(\pi^+ \, \pi^-)$ system. The interference between different 
partial waves could give rise to a nonzero chiral-odd fragmentation function 
called the interference fragmentation function (IFF). 
The IFF would lead to single spin asymmetries in 
$ e \, p^\uparrow \rightarrow e' \, (\pi^+ \, \pi^-) \, X$ and 
$p\, p^\uparrow \to (\pi^+ \, \pi^-) \, X$, both proportional to the
transversity function. The SSA expression for 
$e \, p^\uparrow \rightarrow e' \, (\pi^+ \, \pi^-) \, X$ is \cite{JJT} 
\beq
\big\langle \cos( \phi_{S_T}^e + \phi_{R_T}^e) 
\big\rangle \propto 
\vert \bm S_{T}\vert \vert 
\bm R_{T}\vert \, F(m^2) \, \sum_{a,\bar a} e_a^2
\,x\,h_{1}^{a}(x) \delta\hat q_I^{}(z), 
\eeq
where $z=z^+ + z^-$; $\bm R_{T} 
= (z^+ \bm k^- - z^- \bm k^+)/z$; $m^2$ is the $\pi^+ \pi^-$ invariant 
mass; and $F(m^2) = \sin \delta_0 \sin \delta_1
\sin(\delta_0-\delta_1)$, where $\delta_0, \delta_1$ are the $\ell = 0, 1$ 
phase shifts.
Note the implicit assumption of factorization of $z$ and $m^2$ 
dependence, which leads to the prediction that on the $\rho$
resonance the asymmetry is zero (according to the
experimentally determined phase shifts). But more general $z, m^2$ 
dependences have been considered \cite{Bianconi} and should be tested. 

The asymmetry expression is based on a collinear factorization 
theorem (soft gluon contributions cancel, no Sudakov factors appear). 
Theoretically this is very clean and an analysis beyond tree level is
conceptually straightforward. 
The evolution of $\delta \hat{q}_I(z)$ is the same as for the transversity
{\em fragmentation\/} function
$H_1(z)$, e.g.\ the LO evolution equation is given by 
\beq
\frac{\partial \; z \delta \hat{q}_I(z)}{\partial \ln Q^2} = 
\frac{\alpha_s(Q^2)}{2 \pi} C_F \int_z^1 dy \; 
\left[\frac{3}{2}\,\delta(y-z) + 
\frac{2z}{y(y-z)_+}\right] y \delta \hat{q}_I(y).
\eeq
A NLO analysis is feasible and analogous to $A_{TT}^{DY} \propto  
\cos( \phi_{S_{1T}}^e + \phi_{S_{2T}}^e)h_1 \bar{h}_1$. 

For the extraction of the interference fragmentation functions one can study a 
$\cos(\phi_{R_{1T}}^e + \phi_{R_{2T}}^e)$ asymmetry \cite{Artru-Collins} in   
$e^+ \, e^- \, \to \, (\pi^+ \, \pi^-)_{\text{jet} \, 1} \, (\pi^+ \,
\pi^-)_{\text{jet} \, 2} \, X$ which is proportional to $(\delta\hat
q_I)^{2}$. This is again possible at BELLE and in this case there is no 
expected asymmetric background. Combining such a result with for instance  
the single spin asymmetry in $p\, p^\uparrow \to \pi^+ \pi^- \, X$ to be 
measured at RHIC, seems --at present-- to be one of the
most realistic ways of obtaining information on the transversity function. 

\section*{Acknowledgments}

\noindent
I thank Matthias Grosse Perdekamp, Bob Jaffe, Jens Soeren Lange, Akio
\mbox{Ogawa}, 
Werner Vogelsang for helpful discussions. Furthermore, I thank RIKEN, 
Brookhaven National Laboratory and 
the U.S.\ Department of Energy (contract number DE-AC02-98CH10886) for
providing the facilities essential for the completion of this work.

\end{document}